\documentclass[prb,twocolumn,showpacs]{revtex4}
\usepackage{amsmath}
\usepackage{amssymb}
\usepackage{bm}
\usepackage{epsfig}
\usepackage{graphicx}
\usepackage{color}

\renewcommand{\Re}{\mathop{\rm Re}}
\renewcommand{\Im}{\mathop{\rm Im}}

\begin{document}

\title{Effect of pump-probe detuning on the Faraday rotation and ellipticity signals of mode-locked spins in InGaAs quantum dots}
\author{M.~M.~Glazov$^{1}$, I.~A.~Yugova$^{2}$, S.~Spatzek$^{3}$, A.~Schwan$^{3}$, S.~Varwig$^{3}$, D.~R. Yakovlev$^{1,3}$, D.~Reuter$^4$, A.~D. Wieck$^4$, and M. Bayer$^3$} \affiliation{$^1$ A.~F. Ioffe
Physical-Technical Institute, Russian Academy of Sciences, 194021
St.~Petersburg, Russia} \affiliation{$^2$ Physical Faculty of St.Petersburg State University,
198504 St.~Petersburg, Russia} \affiliation{$^3$
Experimentelle Physik 2, Technische Universit\"at Dortmund,
D-44221 Dortmund, Germany} \affiliation{$^4$ Angewandte
Festk\"orperphysik, Ruhr-Universit\"at Bochum, D-44780 Bochum,
Germany}

\begin{abstract}
We have studied the Faraday rotation and ellipticity signals in
ensembles of singly-charged (In,Ga)As/GaAs quantum dots by pump-probe spectroscopy. For
degenerate pump and probe we observe that the Faraday rotation
signal amplitude first grows with increasing the time separation between pump
and probe before a decay is observed for large temporal separations.
The temporal behavior of the ellipticity signal, on the other hand, is regular:
its amplitude decays with the separation. By contrast, for detuned pump
and probe the Faraday rotation and ellipticty signals both exhibit
similar and conventional behavior.
The experimental results are well described in the frame of a
recently developed microscopic theory [Phys. Rev. B {\bf 80}, 104436
(2009)]. The comparison between calculations and experimental data
allows us to provide insight into the spectral dependence of the
electron spin precession frequencies and extract the electron $g$
factor dependence on energy.
\end{abstract}
\date{\today}
\maketitle

\section{Introduction}

Studies of electron spin dynamics in solids have recently become a
rapidly developing field of condensed matter
physics.\cite{Spintronics_02,Spintronics_08} Insight into the
electron spin dynamics is provided by the pump-probe technique which
has proven to be a versatile tool to address spin
coherence.\cite{Beschoten05} The basic idea behind this method is
quite simple: the sample is illuminated by the circularly
polarized pump pulse which creates a non-equilibrium spin
orientation of charge carriers. After a certain time delay the
weaker linearly polarized probe pulse hits the sample. The rotation
of its polarization plane in transmission (spin Faraday effect)  or
in reflection (spin Kerr effect) geometry as well as the degree of
its ellipticity serve as measures of the carrier spin polarization. Applied
to semiconductor systems, the pump-probe technique allowed one to
measure spin relaxation times, image electron spin diffusion,
monitor spin precession caused by an external magnetic field and
spin-orbit coupling, etc [see
Refs.~\onlinecite{Spintronics_08,yugova09} and references therein].

The pump-probe technique has also turned out to be extremely useful
to study the specifics of spin dynamics in $n$-type singly-charged
semiconductor quantum dots (QDs) where long-lived spin coherence of
resident electrons can be efficiently
generated.\cite{Greilich_PRL06} The pump-probe method revealed the
effects of spin mode-locking \cite{Greilich_Science06} and of the
coupled electron-nuclear spin dynamics.\cite{greilich07}

While the pump-probe technique is widely used experimentally, the
microscopic processes responsible for Faraday/Kerr rotation and
ellipticity signals were established for $n$-type QD systems only
recently.\cite{yugova09} In particular, it was demonstrated that, the Faraday
rotation (FR) and ellipticity signals have strongly different spectral
dependencies and, in case of a QD array, can provide information
about different subensembles of electrons. The developed theory
suggested that the temporal evolution of these signals can be quite
different and depends strongly on the energy detuning between the
pump and probe pulses. Although in transverse magnetic field both
ellipticity and Faraday rotation demonstrate oscillations resulting
from the spin precession, for degenerate pump and probe pulses, the
Faraday rotation signal amplitude may grow with an increase of the
time separation between pump and probe pulses, contrary to the decay of the
ellipticity signal. The observed spin beats frequencies are also
different for the Faraday rotation and ellipticity signals.

This paper aims at illustrating experimentally the theoretical
predictions of Ref.~\onlinecite{yugova09}. We present experimental
data on Faraday rotation and ellipticity signals in $n$-type
(In,Ga)As/GaAs singly-charged QDs in the vicinity of the trion
resonance, where the pump and probe energies have been varied relative
to each other. This particular experimental configuration (i)
enables us to monitor the spin dynamics of resident carriers, and
(ii) allows a direct comparison with the theory of
Ref.~\onlinecite{yugova09}. We demonstrate good agreement of the
experimental data with the theory.

In Section~\ref{Sec:exp} we present an overview of the studied
sample and the experimental techniques. The experimental results are
summarized in Sec.~\ref{Sec:results}. Section~\ref{Sec:theory}
discusses the theoretical background and presents the comparison of
the experimental data with the model calculations. A brief
summary of the results is given in Sec.~\ref{Sec:concl}.

\section{Sample and experimental technique}
\label{Sec:exp}

The studied heterostructure was grown by molecular-beam epitaxy and
consists of 20 layers of (In,Ga)As/GaAs QDs separated by 60~nm GaAs
barriers with a QD density of about $10^{10}$~cm$^2$ in each layer.\cite{Greilich_Science06,yu07} The $\delta$-sheets of Si donors were positioned at the distance of
20 nm below each QD layer with the dopant density being roughly
equal to the dot density in order to achieve an average occupation
of a single electron per dot. The as-grown InAs/GaAs sample shows ground-state
photoluminescence around $1.05$~eV at cryogenic temperatures. It was thermally
annealed for 30 seconds at 945$^{\mathrm o}$C to shift the emission
energy to 1.398~eV due to Ga diffusion into InAs QDs. 

In the experiments the spin coherence was measured by a
time-resolved two-color pump-probe technique. Two pulsed Ti:Sapphire
lasers acting as pump and probe were synchronized to each other at a
repetition rate of 75.75 MHz corresponding to a repetition period of
$T_R=13.2$~ns. The pulse duration of both lasers was 2 ps, and their
photon energies could be tuned independently. The power ratio of the lasers
was about 5 (pump) to 1 (probe). The pump beam was modulated with a
photoelastic modulator, changing between left- and
right-handed circular polarizations at 50 kHz frequency. The
probe beam was linearly polarized.  The pump laser energy was tuned
in resonance with the maximum of the photoluminescence of the QDs,
$\hbar \omega_{\rm P}=1.398$~meV (see Refs.
\onlinecite{Greilich_PRL06,Greilich_Science06} for details), while
the probe energy was varied relative to the pump one. We
used a lock-in amplifier technique for time-resolved Faraday
rotation and ellipticity measurements to monitor the
spin coherence excited by the circularly polarized pump. The time
delay between pump and probe was tuned by a mechanical delay line by
which delays up to 6.6 ns with a precision of 7 fs could be scanned.
After passage through the sample the probe beam was sent through a
$\lambda/2$ (for Faraday rotation) or $\lambda/4$ (for ellipticity)
plates and the intensities of the two contained orthogonal
polarizations were measured by a balanced photodiode bridge.

The sample was mounted in a cryostat with a split-coil superconducting magnet and cooled down
to a temperature $T=6$~K. An external magnetic field was applied in
the plane of the sample orthogonal to the light propagation
direction (Voigt geometry). Its strength was chosen to be $B=4$~T.

\section{Experimental results}
\label{Sec:results}

The important property of the QD samples under study, which is
crucial for the FR and ellipticity signal formation, is their
inhomogeneity. Indeed, photoluminescence reveals a broad ($\sim
10$~meV) distribution of the singlet trion resonance frequencies in the
sample.\cite{Greilich_PRL06} The pump pulse excites only a part of this distribution
according to its spectral width of about $1$~meV.  The excited
subensemble contains around a million QDs with different trion resonance
frequencies, $\omega_0$, and different spin precession (Larmor) frequencies
$\Omega_{\rm L}$. The interaction of a circularly polarized photon
with a quantum dot electron is spin-dependent, and therefore can
result in generation of spin coherence in the
QD.\cite{Greilich_PRL06}

The spin dynamics is probed by the weak linearly polarized probe
pulse. The response to this pulse is also dominated by the trion
formation: depending on the electron spin orientation one of the
circular components of the linear pulse interacts with the quantum
dot more efficiently than the other one. As a result the probe pulse
transmitted through the sample acquires a degree of ellipticity, and
its polarization plane is rotated as compared with its initial
orientation.

\begin{figure}[htbp]
\includegraphics[width=\linewidth]{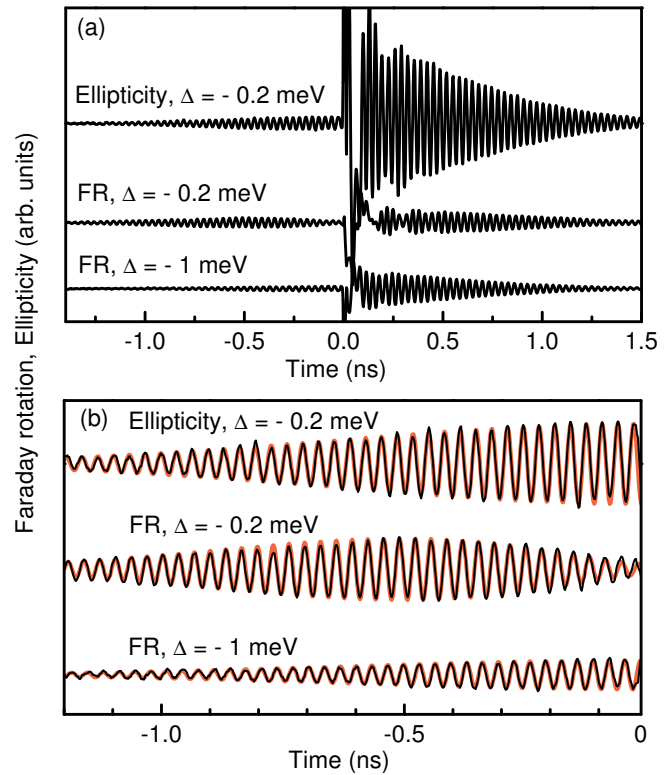}
\caption{(Color online)~(a) Time-resolved Faraday rotation and
ellipticity signals.  Top two curves show ellipticity and Faraday
rotation signals for almost degenerate pump and probe ($\hbar\omega_{\rm P} - \hbar \omega_{\rm pr} =\Delta=-0.2$~meV), bottom
curve presents the Faraday rotation signal for detuned probe
($\Delta=-1.0$~meV). Panel (b) shows close-ups of the corresponding
signals at negative time delay. Thin black curves are the
experimental data, thick red curves are the results of fitting.
$T=6$~K, $B=4$~T.} \label{fig:exp1}
\end{figure}

Experimental ellipticity and Faraday rotation signals as functions
of the pump-probe time delay are presented in Fig.~\ref{fig:exp1}(a).
The top two curves show signals for almost degenerate pump and
probe. The small detuning between pump, $\hbar \omega_{\rm P}$, and
probe, $\hbar \omega_{\rm pr}$, energies, $\Delta = \hbar
\omega_{\rm P}-\hbar \omega_{\rm pr}= -0.2$~meV, was chosen to
compensate a slight asymmetry of the photoexcited QD distribution
and to make the features of the FR signal more pronounced, see
below.\cite{note1} The lowest curve shows FR signal for strongly
detuned pump and probe ($\Delta=-1.0$~meV). Parts of these signals
at negative delays are magnified in Fig.~\ref{fig:exp1}(b). The red
curves in the figure show fits of the data according to
\begin{multline}
\label{signal}
\mathcal E(t), \mathcal F(t)= \sum_i\left[\alpha_i \cos{(\Omega_i | t| - \varphi_i)} +\right. \\
\left. \beta_i| t| \sin{(\Omega_i | t| - \varphi_i)}\right]\exp{\left(-\frac{ t^2}{2\tau_i^2}\right)},
\end{multline}
where $\mathcal E(t)$, $\mathcal F(t)$ are the ellipticity or FR
signals as function of the delay $t$ between the pump pulse and the
subsequent probe pulse.  The subscript $i$ enumerates different
components in the experimental signal related with the spin dynamics of different carriers and their complexes, see below, $\alpha_i$ and $\beta_i$ are the signal amplitudes, $\tau_i$ is the decay
time and $\varphi_i$ is the initial phase of the oscillatory
components. The signals demonstrate oscillations resulting from the
spin precession of electrons and holes about the in-plane magnetic
field.\cite{Greilich_PRL06,yu07} The temporal dependence of the envelope function is
qualitatively different for the FR and ellipticity signals in the
(almost) degenerate pump-probe case. The envelope of the ellipticity
signal decays with an increase of the time separation $|t|$, while
the FR signal first grows with time separation before a decay is
observed. This is especially well manifested at negative delays,
see middle curve in Fig.~\ref{fig:exp1}(b). The increase of FR signal
with time separation $|t|$ is an important feature of the experimental data observed
at small pump-probe detunings $|\Delta|\lesssim 0.5$~meV.

For strongly detuned pump and probe the increase of FR rotation
signal with pump-probe time separation is absent [lowest curve in
Fig.~\ref{fig:exp1}(b)], and the FR signal becomes similar to the
ellipticity one for the same detuning (not shown).

\begin{figure}[htbp]
\includegraphics[width=\linewidth]{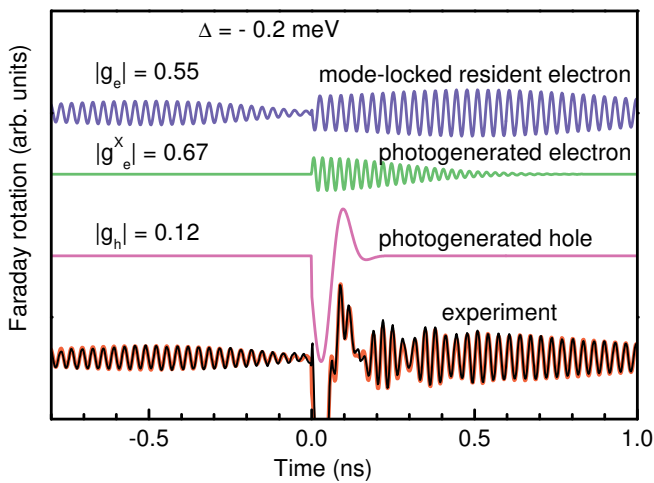}
\caption{(Color online) Faraday rotation signal (lowest curves) for
almost degenerate pump and probe pulses: experimental data shown by
the black curve are superimposed on the fit (red curve). Three top
dependencies are components to the fit (from top to bottom): signal
due to long-lived electron spin polarization in charged QDs, signal
due to the electron-in-exciton spin precession and signal due to
hole-in-exciton spin precession in neutral QDs.\cite{yu07} $T=6$~K,
$B=4$~T. } \label{fig:fit}
\end{figure}

The complicated shape of the observed signals demonstrates that
different physical processes are involved in the formation of the
signals. In order to gain insight into the effects
responsible for the signals and to isolate the contribution caused
by the resident electrons we fitted the experimental data using
Eq.~\eqref{signal}. The different components contained in the
experimental signal are demonstrated in Fig.~\ref{fig:fit}. The
lowest black curve represents the experimental data. It is
superimposed on the red fit curve, which consists of three components
that are shown separately at the top of Fig.~\ref{fig:fit}. These
three dependencies are labeled by the $g$ factor values
obtained from the spin beats frequency by $\hbar \Omega_L = g \mu_B
B$, where $\Omega_L$ is the beats frequency and $\mu_B$ is the Bohr
magneton. Two of these three components appear only at positive
delays. They are attributed to the spin precession of the hole
($|g_{h}|=0.12$) and the electron forming the exciton in neutral
QDs ($|g^{X}_{e}|=0.67$). This conclusion is supported by the
values of the $g$ factors and the decay times being in agreement
with Ref.~\onlinecite{yu07}. Note, that the effective $g$ factor of
the electron-in-exciton component $|g^{X}_{e}|=0.67$ includes the
contribution of the electron-hole exchange interaction, and
therefore it differs from the resident electron $g$ factor at the
same transition energy. 

The main part of the signal, shown by the
top (blue) curve in Fig.~\ref{fig:fit}, is due to the long-lived
electron spin coherence in charged QDs: its lifetime exceeds by far
the radiative lifetimes of excitons and trions in these quantum
dots\cite{yu07} and the extracted $g$ factor values are consistent
with those of resident electrons with the same transition energy.
The almost exact agreement between the fit and the experiment is
demonstrated [red curve in the bottom of Fig.~\ref{fig:fit} determined using
Eq.~\eqref{signal}] suggesting that all main contributions to the
measured FR signal are taken into account.

\begin{figure}[htbp]
\includegraphics[width=\linewidth]{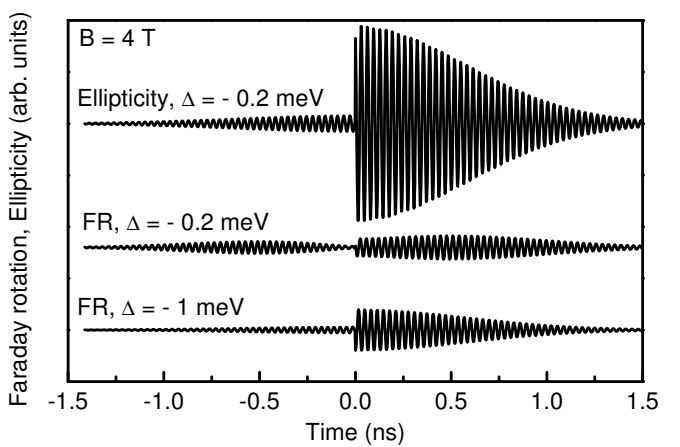}
\caption{Long-lived parts of the Faraday rotation and ellipticity
signals shown in Fig.~\ref{fig:exp1}(a) as extracted from the fit
(shown in Fig.~\ref{fig:fit}). } \label{fig:long_part}
\end{figure}

The described fitting procedure was applied to all measured curves.
Figure~\ref{fig:long_part} displays the extracted long-lived
electron spin coherence contributions for the ellipticity and FR
signals presented Fig.~\ref{fig:exp1}(a). It is clearly seen that for
the degenerate pump-probe conditions ($\Delta = -0.2$~meV) the FR signal increases with
increasing pump-probe time separation, $|t|$. This behavior is in
striking contrast to the ellipticity signal and the FR signal for
detuned pump and probe.

\begin{figure}[htbp]
\includegraphics[width=\linewidth]{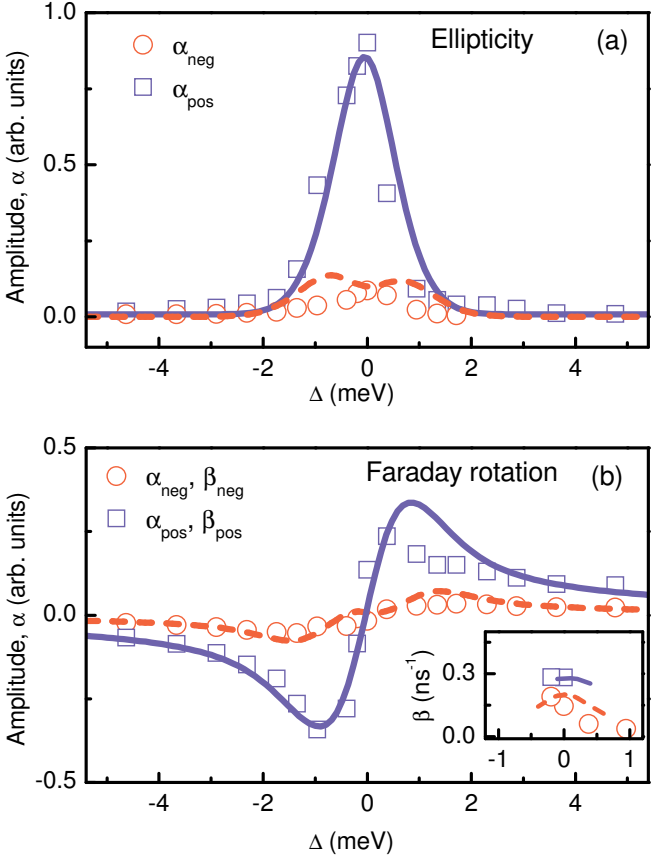}
\caption{(Color online) Amplitudes of the ellipticity signal (a) and
Faraday rotation signal (b) as function of pump-probe detuning.
Circles show the amplitudes $\alpha_{\rm neg}$ of the decaying
component of the signal at negative delays and squares show the
corresponding amplitudes $\alpha_{\rm pos}$ at positive delays, see
Eq.~\eqref{signal}. Inset in panel (b): amplitudes of the growing
component of the FR signal $\beta_{\rm neg}$ (circles) at negative
delays and $\beta_{\rm pos}$ (squares) at positive delays. Lines are
theoretical calculations in the frame of the model developed in
Ref.~[\onlinecite{yugova09}], see Sec.~\ref{Sec:theory}.}
\label{fig:amp1}
\end{figure}

For an in-depth analysis we plot in Fig.~\ref{fig:amp1} the
amplitudes of the long-lived components of the ellipticity and FR
signals extracted from the fit to the experimental data as functions
of pump-probe detuning. For the ellipticity signal the contribution
to the signal amplitude that is growing with pump-probe time separation
$|t|$ is not observed, i.e. $\beta_{\rm neg}=\beta_{\rm pos}=0$.
Therefore, we plotted only the amplitudes $\alpha_{\rm neg}$ for
negative (circles) and $\alpha_{\rm pos}$ for positive (squares)
delays. For the FR signal both the decaying- and the
growing-with-separation contributions are substantial: the
amplitudes $\alpha$ of the decaying component are shown in the main
panel of Fig.~\ref{fig:amp1}(b), and the amplitudes $\beta$ of the
growing component are given in the inset for positive (squares) and
negative (circles) time delays.

One sees in Fig.~\ref{fig:amp1}(a)
that the amplitude of the ellipticity signal is maximum for almost
degenerate pump and probe, while the amplitude of the decaying part
of the FR signal is (within the accuracy of the
measurements\cite{note1}) an odd function of the detuning. Its
absolute value reaches a maximum for pulses detuned by about 1~meV,
see Fig.~\ref{fig:amp1}(b), in agreement with previous studies on
quantum well~\cite{fokina10} and quantum
dot~\cite{mikkelsen,carter09} samples. On the other hand, the
component of the FR signal that is growing with time separation
has not been observed in previous studies to the best of our
knowledge. Its spectral behavior is opposite to that of the decaying part of the FR signal, as it reaches
maximum for degenerate pump and probe and rapidly decreases with
detuning, as shown in the inset of Fig.~\ref{fig:amp1}(b).

\begin{figure}[htbp]
\includegraphics[width=\linewidth]{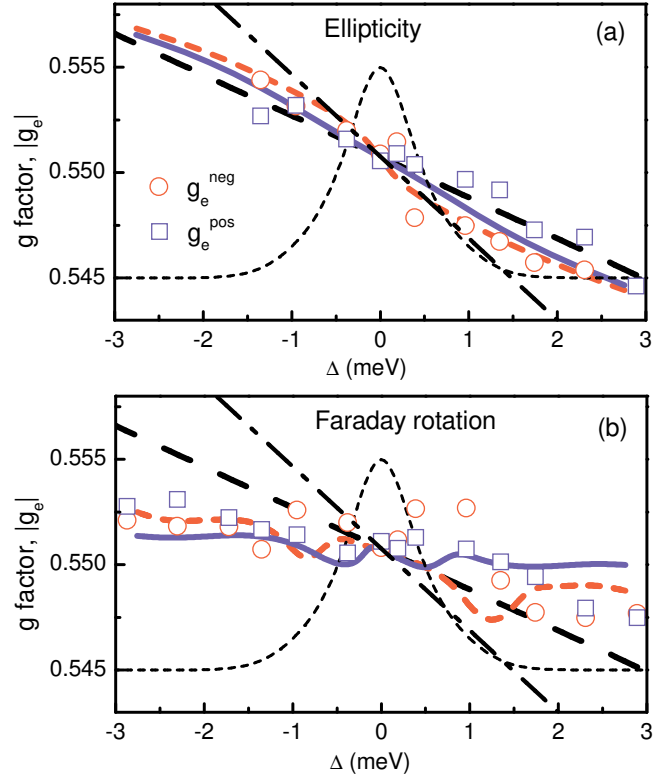}
\caption{(Color online) Absolute values of electron $g$ factors extracted from
ellipticity signals (a) and Faraday rotation signals (b) as
functions of pump-probe detuning. Circles and squares are
experimental data obtained at negative and positive time delays,
respectively. Dash-dotted line presents spectral dependence of
electron $g$ factor calculated after Eq.~\eqref{g:omega0} with the
constants $a=-0.004$~meV$^{-1}$ and 
$c=6.142$ 
taken from fitting the ellipticity by
Eq.~\eqref{ell1} as function of $\Delta$.
Dashed line shows
$|g_e|$ dependence calculated in the frame of a simple analytical
model, Eq.~\eqref{g:Delta}.  Red (dashed) and blue (solid) curves
are the results of theoretical calculations (as described in
Sec.~\ref{Sec:num}). Short dashed curve is integrated over the
additional spread of Larmor frequencies, $\Omega_{\rm L}$, $S_z^+$
[see Eqs.~\eqref{signals} and \eqref{rho:g}].} \label{fig:gfact}
\end{figure}

Finally, let us consider the dependence of the spin beats frequencies
on the pump-probe detuning measured for the ellipticity and FR
signals. It is well established that the electron $g$ factor value
and, correspondingly, the Larmor frequency have energy dispersion
due to the effective band gap variation and, hence, they depend on
the trion resonance (optical transition) frequency
$\omega_0$.\cite{ivchenko05a,Greilich_Science06,kiselev07} Over
narrow energy ranges such a dependence can be well approximated by a
linear function as\cite{yugova09}
\begin{equation} \label{g:omega0}
|g_e(\omega_0)| = a \hbar\omega_0 + c\:,
\end{equation}
with $a$ and $c$ being constants. Figure~\ref{fig:gfact} shows the
detuning dependencies of $|g_e|$ extracted from the ellipticity signal
[panel (a)] and the Faraday rotation signal [panel (b)]. Circles
show the values of $g^{\rm neg}_e$, measured at negative delays and
squares show the $g$-factors, $g^{\rm pos}_e$, at positive delays. For the FR signal the $g$ factors corresponding to the
growing and decaying components are identical, therefore we plot
only values corresponding to the decaying part of the signal.

A good correlation of the $g$-factor values obtained at positive and
negative delays is seen. It indicates that the same electron
subensembles contribute to the signals at positive and negative
delays. Although the $g$ factors extracted from ellipticity and FR
signals are quite close to each other, one can see from
Fig.~\ref{fig:gfact}, that the spectral dependencies of $g$ factor
from the ellipticity signals have a different slope as those from
the FR signals. The FR $g$ factor data shown in panel (b) has a much
weaker spectral dependence then the ellipticity data, shown in panel
(a). Moreover, the spectral dependence of the $g$ factor values
extracted from ellipticity is weaker than the one predicted by
Eq.~\eqref{g:omega0}, dash-dotted line in Fig.~\ref{fig:gfact}(a).

The main experimental findings can be summarized as follows:
\begin{itemize}
\item Faraday rotation and ellipticity signals measured for degenerate pump-probe conditions have drastically different temporal dependencies: the ellipticity signal amplitude decays with time separation between pump and probe, while the Faraday rotation signal contains, in addition to the decaying part, a growing component.
\item For relatively large (as compared with the pump and probe pulse spectral widths) detunings the growing component of Faraday rotation signal disappears so that FR becomes similar to the ellipticity signal.
\item The amplitude of the decaying component of ellipticity signal is an even function of the pump-probe detuning, while that of the Faraday rotation signal is an odd function.
\item The dependencies of the electron $g$ factor on probe spectral position, extracted from the ellipticity and Faraday rotation measurements, are different: the Faraday rotation $g$ factor values demonstrate a weak spectral dependence in contrast to the ellipticity ones.
\item The slope of the electron $g$-factor dependence on the probe spectral position is smaller than one in Eq.~\eqref{g:omega0}.
\end{itemize}
Below we present qualitative and quantitative discussion of these
features.

\section{Theory and discussion}
\label{Sec:theory}
\subsection{General considerations}\label{Sec:theor:gen}

In order to model the temporal dynamics of the Faraday rotation and
ellipticity signals we follow the methods developed in
Ref.~\onlinecite{yugova09}. We consider an array of singly-charged
$n$-type QDs subject to a train of circularly polarized pump pulses
with the optical transition frequency $\omega_{\rm P}$. It is
assumed that $\omega_{\rm P}$ is close to the trion resonance
frequency, $\omega_0$, therefore only one optical transition related
with the formation of the singlet trion state is relevant. Depending
on the initial spin orientation of the resident electron before pump
pulse arrival, $\bm S^{(b)}$, trion formation is either possible or
blocked due to the Pauli principle for a given circular polarization
of the pump pulse. As a result, electrons with a certain spin
$z$-component, i.e. $-1/2$, take part in the trion formation for
$\sigma^+$ circularly polarized light and they are left
depolarized after trion recombination due to the fast hole spin
relaxation~\cite{zhu07,yugova09} (see also
Ref.~\onlinecite{fokina10} for other regimes of electron spin
coherence initialization). Therefore the electron spin before, $\bm
S^{(b)}$, and after, $\bm S^{(a)}$, the pump pulse are correlated:
there is an additive contribution which describes spin coherence
generation by the pump pulse and a non-additive one which depends on
(i) the pump pulse parameters and (ii) $\bm S^{(b)}$ and describes spin
change by the pump pulse action.\cite{yugova09,economou06,zhu10}
Explicit expressions for this action are presented by Eqs. (16) in
Ref.~\onlinecite{yugova09}.

Between the pump pulses the resident electron spins precess about
the in-plane external field $\bm B$ and decay due to spin relaxation
processes. The latter can be characterized by a single time constant
$\tau_s$. Usually, $\tau_s$ exceeds by far the pulse repetition
period, $T_R$, so that a steady state distribution of precessing
spins is formed.~\cite{Greilich_Science06,yugova09} We denote by $S_z^{+}(\omega_0;\omega_{\rm P})$ the
electron spin $z$ component in the steady state right after the arrival of the pump
pulse in a QD with the trion resonance frequency
$\omega_0$.

The time-resolved ellipticity $\mathcal E(t)$ and Faraday rotation
$\mathcal F(t)$ signals measured by the probe with optical frequency
$\omega_{\rm pr}$ are given by the real and imaginary parts of the
following convolution~\cite{yugova09}
\begin{multline}
\label{signals}
\mathcal E(t) + \mathrm i \mathcal F(t) = \int p(\omega_0,\Omega_{\rm L}) G(\omega_{\rm pr} - \omega_0) S_z^{+}(\omega_0,\omega_{\rm P}) \times \\
\cos{[\Omega_{\rm L} t + \varphi(\omega_0, \Omega_{\rm L})]} \exp(- t/\tau_s) \mathrm d\omega_0 \mathrm d\Omega_{\rm L}.
\end{multline}
Here the delay between the pump pulse and the next probe pulse
$t>0$. Function $p(\omega_0,\Omega_{\rm L})$ is the joint distribution
function of optical and Larmor frequencies of QDs, and the function
$G(\Lambda)$ characterizes the spectral sensitivity of the
ellipticty and FR signals with $\Lambda$ being the detuning between
the probe carrier frequency and the QD trion resonance frequency. The last
two factors in Eq.~\eqref{signals} describe the dynamics of a single
spin in a given QD, which includes spin precession and spin
relaxation. Here, $\varphi(\omega_0,\Omega_{\rm L})$ is the initial
spin phase.\cite{yugova09,zhu07}

\subsection{Illustrative example}\label{Sec:example}

It is the combined effect of the spectral variation of the function
$G(\Lambda)$ and the dependence of the Larmor frequency on the
optical frequency which determines the temporal evolution of the
ellipticity and FR signals. In order to illustrate the qualitative
behavior of these signals we employ a simple model, in which the
function $G(\Lambda)$ is assumed to have the form
\begin{equation}
\label{Glambda}
G(\Lambda) =  (1+ 2 \mathrm i \Lambda \tau_p)\exp{[-(\Lambda\tau_p)^2]},
\end{equation}
where $\tau_p$ is the pulse duration and $\Lambda=\omega_{\rm pr} -
\omega_0$ is the detuning between the probe pulse and the quantum
dot trion resonance frequency. This function captures the main features
of ellipticity and FR spectral sensitivity: the ellipticity has its
maximum for $\Lambda=0$, while the Faraday rotation is an odd
function of the detuning, having maxima of opposite signs on its
wings. The form of $G(\Lambda)$ is approximate, nevertheless for
$\Lambda\tau_p\lesssim 1$ the function $G(\Lambda)$ given by
Eq.~\eqref{Glambda}, is quite similar to the behavior of the exact
spectral function for Rosen\&Zener pulses, see Eq.~(61) in
Ref.~\onlinecite{yugova09}. It is worth to mention that for
$\Lambda\tau_p \gg 1$ the imaginary part of $G$ (i.e. the FR
sensitivity) decays faster than the exact function,\cite{note2} but
this does not affect the temporal behavior of the signals on a
qualitative level.

For the distribution function of the optical and Larmor precession
frequencies, $p(\omega_0, \Omega_{\rm L})$, we assume that: (i) the
distribution of the optical frequencies is rather smooth, and (ii)
there is a correlation between the electron $g$ factor and the trion
resonance frequency given by Eq.~\eqref{g:omega0}:
\begin{equation} \label{rho:g}
p(\omega_0,\Omega_{\rm L})=\frac{1}{{\sqrt{2\pi} \Delta \Omega }} \exp\left[ - \frac{(\Omega_{\rm L} - \mu_B g_e({\omega}_0)B/\hbar
)^2}{2 (\Delta \Omega)^2}\right],
\end{equation}
where $\Delta \Omega $ is the the spin precession frequency
dispersion related, e.g., to nuclear field fluctuations. Finally, we
assume that the steady state distribution of precessing spins is described by a Gaussian (the effects of mode-locking which result in the modification of $S_z^+$ for certain Larmor and optical frequencies\cite{yugova09} are discussed below)
\begin{equation}
\label{sz0}
S_z^{+}(\omega_0, \omega_{\rm P}) = S_0\rm \exp{[-(\omega_0 - \omega_{\rm P})^2\tau_p^2]},
\end{equation}
with $S_0$ being a constant, which depends on the pump amplitude.

Under these assumptions the convolutions in Eq.~\eqref{signals} can
be easily evaluated.\cite{note4} To simplify the analysis we
disregard the fluctuations of the spin precession frequency, which
are not correlated with the resonance frequency, i.e., put $\Delta
\Omega=0$ in Eq.~\eqref{rho:g} and additionally consider the limit
of $\tau_s\to \infty$. Inclusion of these effects would simply
enhance the decay of the spin beats. We also ignore the initial
phase of the spin beats [i.e. we put $\varphi(\omega_0,\Omega_{\rm
L})=0$] because it does not bring in any new physics.  The resulting
signals are
\begin{subequations}
\begin{equation}
\label{ell1}
\mathcal E(t) = \sqrt{\frac{\pi}{2\tau_p^2}} \exp{\left[\frac{-\Delta^2\tau_p^2/(2\hbar^2)-(\Omega't)^2}{8\tau_p^2}\right]} \cos{\left(\tilde{\Omega}_0  t\right)} ,
\end{equation}
\begin{multline}
\label{far1}
\mathcal F(t) =
\frac{1}{2}
\sqrt{\frac{\pi}{2\tau_p^2}}
 \exp{\left[
 \frac{-\Delta^2\tau_p^2/(2\hbar^2)-(\Omega't)^2}{8\tau_p^2}
 \right]}
 \times\\
\left[\frac{2\Delta\tau_p}{\hbar}
\cos{\left(\tilde{\Omega}_0 t\right)}
+\frac{\Omega' t}{\tau_p}
\sin{\left(\tilde{\Omega}_0 t\right)}
\right].
\end{multline}
\end{subequations}
Here we introduced the following notations: $\Omega'=\mathrm
d\Omega_{\rm L}/\mathrm d\omega_0 = \mu_B a B$,
$\Delta/\hbar = \omega_{\rm P} - \omega_{\rm pr}$ is the pump-probe
detuning and $\tilde{\Omega}_0 = \Omega_0 + \Omega'\Delta/(2\hbar)$
is the effective spin precession frequency, with
$\hbar\Omega_0=g_e(\omega_{ \rm pr})\mu_B B$.

It follows from Eqs.~\eqref{ell1} and \eqref{far1} that the temporal
behavior of the FR and ellipticity signals can be strongly
different. This is especially well seen for degenerate pump and
probe where the amplitude of the FR signal first grows with
pump-probe time separation and afterwards decays, while the ellipticity
signal amplitude simply decays. This turns out to be a direct
consequence of the correlation between Larmor frequency $\Omega_{\rm
L}$ and optical frequency.

\begin{figure}[htbp]
\includegraphics[width=\linewidth]{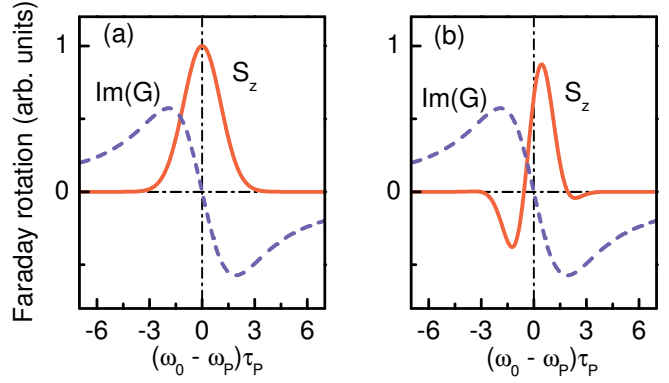}
\caption{(Color online) Schematic illustration of Faraday rotation
signal formation for degenerate pump-probe conditions, $\omega_{\rm pr} = \omega_{\rm P}$. Panel (a) corresponds to zero pump-probe delay,
panel (b) to a pump-probe delay $t>0$. Red curve shows initial spin
distribution generated by the pump pulse, blue dashed curve shows
the spectral sensitivity of the FR signal $\Im G(\omega_0 - \omega_{\rm pr})$.} \label{fig:detect_farad}
\end{figure}

At $t=0$ the spin distribution is symmetric and gives zero FR signal
since it is convoluted with the odd function, $\Im{G(\Lambda)}$, as
shown in Fig.~\ref{fig:detect_farad}(a). A non-zero Faraday signal can appear only due to some asymmetry. As time goes by, the
electron spin distribution becomes asymmetric since spins in QDs
with larger trion resonance frequencies $\omega_0$ precess slower than the
spins of QDs with smaller resonance frequencies because corresponding electron $g$ factors are different, see Eq.~\eqref{g:omega0}. As a result, the
distribution of electron spins as a function of resonance frequency is
no longer symmetric with respect to the probe frequency $\omega_{\rm pr}$, see Fig.~\ref{fig:detect_farad}(b). Such an imbalance
results in the appearance of non-zero Faraday rotation signal at $
t>0$. At relatively large delays
the spin dephasing caused by the nuclear spin fluctuations and the
spread of $g$-factors comes into play and the Faraday rotation
signal amplitude decreases. The detuning between the pump and probe pulses results in the asymmetry of the spin distribution with respect to $\omega_{\rm pr}$ even at $t=0$. Hence, ordinary, decaying with time component of FR signal appears and growing with time component becomes less pronounced in line with experimental observations,
Fig.~\ref{fig:amp1}(b).

For ellipticity, the spectral function $\Re{G(\Lambda)}$ is even and
is sensitive to the average spin $z$ component which oscillates in
time and decays due to the spread of Larmor frequencies.

The outlined model also explains qualitatively the difference
between the FR and the ellipticity signals at negative pump-probe
delays, $t<0$, and the increase of the FR with increase of
pump-probe time separation in this delay range. The signal at $t<0$
appears because the sample is subject to a train of pump
pulses. If the single electron spin relaxation time $\tau_s$ is much
longer than the pulse repetition period, which is the case in our
experiments,\cite{Greilich_Science06,greilich07} each electron
preserves its spin coherence up to the moment of the next pump pulse
arrival. As a result, spin precession mode-locking occurs: for
electrons with Larmor spin precession periods being commensurable
with the pump repetition period the spin is accumulated as the pump
pulses arrive in phase with the spin precession. Hence, the
steady state distribution of precessing spins,
$S_z^+(\omega_0,\omega_{\rm P})$, has sharp maxima for those QDs
where $\Omega_{\rm L}(\omega_0)T_R=2\pi N$, $N$ being an
integer.\cite{Greilich_Science06,yugova09} If only these mode-locked
spins are taken into account, the precession frequencies are
commensurable with the repetition period and the signals are even
functions of the pump-probe delay, $t$. 
The presence of other spin
precession frequencies results in additional contributions to the
spin signals at positive delays, $t>0$, which dephases towards the
moment of the next pump pulse arrival.

It is possible to obtain an analytical result for the ellipticity
and the FR signals for a Gaussian distribution of the quantum dot
resonant frequencies $p(\omega_0)$. In this case, the FR signal
behavior depends on the spectral position of the pump pulse with
respect to the maximum of the QD trion resonance frequency distribution. Indeed, for the absolutely symmetric situation at $t=0$ the FR signal is zero. A shift of $\hbar \omega_{\rm P}$ from the maximum
results in an asymmetry of the photoexcited QD distribution, and,
correspondingly, in a shift of the maximum of ellipticity and the
zero of FR away from zero detuning.~\cite{yugova09} Such an
asymmetry can be caused also by the transition energy variation of
the trion oscillator strength in the QDs ensemble. 
The asymmetry discussed
above does not require the inclusion of nuclear spin effects to
describe the shifts of ellipticity and Faraday rotation signals from
zero detuning, as was suggested in Ref.~\onlinecite{carter09}.

Now we turn to the spin precession frequencies. In our simplified
model one can see that the effective spin precession frequency
$\tilde{\Omega}_0$ in Eqs.~\eqref{ell1} and \eqref{far1} depends
both on the spectral positions of the pump and the probe. It
corresponds to the averaged optical frequency between pump and
probe. Hence, the observed electron $g$ factor which is evaluated from the experimental signals measured for non-degenerate conditions as function of the
pump-probe detuning is approximately given by
\begin{equation} \label{g:Delta}
|g_e(\Delta)| = a \hbar\omega_{\rm P} + c - \frac{a}{2} \Delta\:,
\end{equation}
i.e. its slope is twice smaller as compared to the slope of the
$|g(\omega_0)|$ dependence, Eq.~\eqref{g:omega0}. This is because the
observed $g$ factor is an average of that at the pump and the probe
frequencies. These functions are shown in Fig.~\ref{fig:gfact} by
the dashed [Eq.~\eqref{g:Delta}] and dash-dotted
[Eq.~\eqref{g:omega0}] lines with constants $a \approx
-0.004$~meV$^{-1}$ and $|g_e(\omega_{\rm P})|\approx 0.55$. The
spectral dependence of the $g$ factor extracted from the ellipticity
signal is well described within this simplified model, see
Fig.~\ref{fig:gfact}(a).

The simplified model, however, does not describe well the $g$ factor
data obtained from the FR signal. It is seen in
Fig.~\ref{fig:gfact}(b) that the experimental data have much weaker
spectral dependence than the theoretical predictions. This
discrepancy results from the fact that, for $\Lambda \tau_p\gtrsim
1$ the detuning dependence of $\Im{G}(\Lambda)$ used in theory is
much stronger than that in experiment. In fact, for weakly decaying
$\Im{G}(\Lambda)$ and relatively large detuning the convolution determining the Faraday rotation,
Eq.~\eqref{signals}, is controlled by the trion resonance frequencies
corresponding to the maximum of $S_z^+(\omega_0,\omega_{\rm P})$,
i.e. the spin precession frequency is approximately given by
$g(\omega_{\rm P}) \mu_B B/\hbar$. Below, these findings are
confirmed by numerical modeling.

\subsection{Numerical modeling} \label{Sec:num}

The simple model discussed in Sec.~\ref{Sec:example} qualitatively
describes the main features of the experiments. To perform a
quantitative analysis and to eliminate the discrepancies between the
FR spectral dependencies and the simple model we did numerical
calculations in the frame of the model proposed in
Ref.~\onlinecite{yugova09}. The FR and ellipticity signals were
calculated from Eq.~\eqref{signals}, with the spectral function
$G(\Lambda)$ corresponding to the Rosen\&Zener laser pulse of proper
duration, see Eq.~(61) in Ref.~\onlinecite{yugova09}. We also took
realistic distributions of the electron $g$ factors and optical
frequencies: the constants $a$ and $c$ in Eq.~\eqref{g:omega0} were
extracted from the spectral dependence of the $g$ factor in
Fig.~\ref{fig:gfact}(a), see also Eq.~\eqref{g:Delta}, so that the
$g$ factor spread within the pulse spectral width was estimated to
be $\Delta g = 0.002$.  By comparing this value with the $g$ factor
spread extracted from the spin beats decay $\Delta g = 0.005$ we
determine the additional fluctuations of the $g$ factor, and,
correspondingly, of the electron spin precession frequency
$\Delta\Omega$, in Eq.~\eqref{rho:g} caused, e.g., by the nuclear
spin fluctuations.\cite{single_mode}

\begin{figure}[htbp]
\includegraphics[width=1.\linewidth]{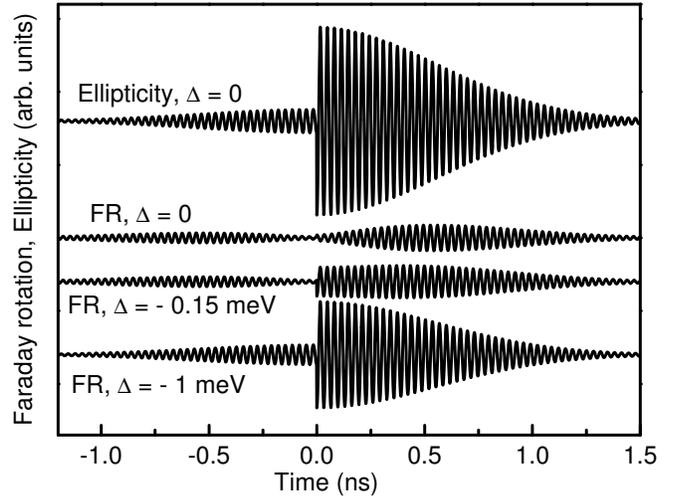}
\caption{Modeled time-resolved Faraday rotation and ellipticity
signals.  The two upper dependencies show ellipticity and Faraday
rotation signals for degenerate pump and probe, the two bottom
dependencies show the Faraday rotation signal for detuned probe. Calculation parameters: $a=-0.004$~meV$^{-1}$, $c=6.142$ 
$\Delta g = 0.005$, $\Theta = 0.2$, $T_R=13.2$~ns, $\tau_p=3$~ps.
}
\label{fig:teor1}
\end{figure}

Figure~\ref{fig:teor1} shows calculated ellipticity and Faraday
rotation signals for degenerate and nondegenerate pump-probe
arrangements. The modeled signals are very similar to the
experimental ones shown in Fig.~\ref{fig:exp1}(a), see also
Fig.~\ref{fig:long_part}. The parameters of calculation were chosen in a such a way that the calculated signals and measured ones are as similar as possible. The effective pulse area $\Theta$, which
determines the electron spin change by the pump pulse,\cite{yugova09}
was estimated from the experimental spin signal amplitudes before and
after the pump pulse arrival to be $\Theta \approx 0.2$. The duration of the Rosen\&Zener pulse used in the calculations was chosen to be $\tau_p=3$~ps in order to reproduce properly the temporal shape of experimental pulse and widths of the signal amplitudes spectral depedencies.

A fitting procedure similar to the one applied to the experimental
data was used also for extraction of the amplitudes of the
calculated ellipticity and Faraday rotation signals. Solid and
dashed lines in Fig.~\ref{fig:amp1} show the calculated spectral
dependencies of these amplitudes, which almost perfectly match the
experimental points. The discrepancy between the calculated and the
experimental $\alpha_{\rm neg}$ for the ellipticity signal in
Fig.~\ref{fig:amp1}(a) may result from nuclear
effects.\cite{greilich07}

The solid and dashed curves in Fig.~\ref{fig:gfact} give the
calculated spectral dependencies of the electron $g$ factor
extracted from the ellipticity and FR signals. The complicated
non-monotonous dependence of the $g$ factor in the FR signal for
moderate detuning results from the complex shape of the spectral
function for the FR signal. The calculation shows, that the
spin beats occur at multiple frequencies corresponding, e.g., to the
maxima and minima of the Faraday rotation spectral function and to
the spin distribution maxima. As a result, fitting of this behavior
by the simplified Eq.~\eqref{signal} gives an averaged $g$ factor
which oscillates with the detuning in the ranges where the spectral
function and the pump-induced spin distribution strongly overlap. In
agreement with the experiment, the spectral dependence of the $g$
factor in the ellipticity signal is stronger than that in FR. Good agreement between the experiment and theory is
achieved.

\section{Conclusions}\label{Sec:concl}

We have studied in detail the Faraday rotation and ellipticity
signals in singly-charged (In,Ga)As/GaAs quantum dot ensembles. Not only the
amplitudes, but also the temporal behavior of the ellipticity and
Faraday rotation signals are qualitatively different. For degenerate
pump and probe pulses the Faraday rotation signal first grows as a
function of time separation between the two pulses and then
decays, while the ellipticity starts from its maximum value for zero
time separation and then decays. We have demonstrated also, that the spin
beat frequencies measured as a function of the probe spectral
position are quite different for Faraday rotation and ellipticity.
In the latter case, the spin precession frequency changes linearly
with pump-probe detuning, while in the former case the spin
precession frequency does not change much with detuning.

The experimental findings are well explained within the developed
theoretical formalism, which takes into account the microscopic
processes responsible for pump-probe signal formation and the
inhomogeneity of the quantum dot ensemble. For degenerate pump and
probe pulses, the Faraday rotation signal is sensitive to the
asymmetry of the electron spin distribution. Such an asymmetry
appears with time due to the correlation between the electron $g$
factor and the trion resonance frequency, $\omega_0$. On the
other hand, the ellipticity signal measures the averaged electron
spin and is therefore much less sensitive to the distribution
asymmetry.

The increase of the Faraday rotation signal with pump-probe
time separation results from the dependence of the electron $g$ factor on
the effective band gap. Hence, a similar temporal behavior can be
observed not only for the long-lived resident electron signal, but
also for the electron spin beats in empty quantum dots. In the
latter case, however, the exchange interaction between electron and
hole in the exciton can make a considerable contribution to the spin
beat frequency, masking the spectral dependence of the $g$ factor.

We have shown both experimentally and theoretically that the Faraday
rotation and ellipticity effects probe different parts of the
inhomogeneous quantum dot ensemble. Quantitative agreement between
theory and experiment, including comparison of the spectral
dependencies of signal amplitudes and spin beats frequencies, has
been achieved.

\acknowledgments We are grateful to Al.L. Efros and E.L. Ivchenko for valuable discussions. This work was supported by the BMBF project
nanoquit, the Deutsche Forschungsgemeinschaft and Russian Foundation for Basic Research. M.M.G. is thankful to the  ``Dynasty''
Foundation---ICFPM  and President grant for young scientists.


\begin{thebibliography}{99}
\bibitem{note1} The maximum of ellipticity signal and the zero of FR signal (also the maxima of amplitudes $\beta$ of the FR signal) are slightly shifted from zero detuning between pump and probe. We discuss possible reasons for this shift below, in Sec.~\ref{Sec:theory}.
\bibitem{note2} Making use of the explicit expressions in Ref.~\onlinecite{yugova09} one can show that $\Im G(\Lambda) \propto (\Lambda\tau_p)^{-1}$.
\bibitem{note4} {It is also possible to obtain analytical expression for exponential pump and probe pulses with envelopes $\propto \exp{(-|t|/\tau_p)}$, but the resulting expressions are cumbersome and do not result in qualitatively new physical effects.}
\end{thebibliography}
\end{document}